\documentclass[prd,12pt,tightenlines,onecolumn,amssymb,preprintnumbers,nofootinbib,superscriptaddress]{revtex4-1}

\pdfoutput=1

\usepackage{hyperref}
\usepackage{feynmf}
\usepackage{graphicx}
\usepackage{enumitem}
\usepackage{latexsym}
\usepackage{amsfonts}
\usepackage{amssymb}
\usepackage{mathrsfs}
\usepackage{color}
\usepackage{amsmath}
\usepackage{slashed}
\usepackage{dcolumn}
\usepackage{verbatim}
\usepackage{comment}
\usepackage{float}
\usepackage{multirow}
\usepackage{xspace}
\usepackage[normalem]{ulem}
\graphicspath{{./}{./Figs/}}

\allowdisplaybreaks

\newcommand{\beq}{\begin{eqnarray}}
\newcommand{\eeq}{\end{eqnarray}}
\newcommand{\bea}{\begin{eqnarray}}
\newcommand{\eea}{\end{eqnarray}}

\newcommand{\gev}{\, {\rm GeV}}
\newcommand{\mev}{\, {\rm MeV}}
\newcommand{\gsim}{\lower.7ex\hbox{$\;\stackrel{\textstyle>}{\sim}\;$}}
\newcommand{\lsim}{\lower.7ex\hbox{$\;\stackrel{\textstyle<}{\sim}\;$}}
\newcommand{\mpl}{M_{\rm Pl}}

\newcommand{\nnmb}{\nonumber}
\newcommand{\del}{\partial}

\newcommand{\lrf}[2]{\left(\frac{#1}{#2}\right)}

\newcommand{\lag}{\mathscr{L}}

\newcommand{\zpp}{0^{++}}
\newcommand{\opm}{1^{+-}}

\newcommand{\jpc}{J^{PC}}
\newcommand{\sun}{SU(N)}
\newcommand{\sunge}{SU(N\geq 3)}
\newcommand{\suthree}{SU(3)}

\begin{document}

\title{Dark Matter from Dark Glueball Dominance
}

\author{David McKeen}
\email{mckeen@triumf.ca}
\affiliation{TRIUMF, 4004 Wesbrook Mall, Vancouver, BC V6T 2A3, Canada}

\author{Riku Mizuta}
\email{rmizuta@triumf.ca}
\affiliation{TRIUMF, 4004 Wesbrook Mall, Vancouver, BC V6T 2A3, Canada}
\affiliation{Department of Physics and Astronomy, University of British Columbia,
6224 Agricultural Road, Vancouver, B.C. V6T 1Z1, Canada}

\author{David E. Morrissey}
\email{dmorri@triumf.ca}
\affiliation{TRIUMF, 4004 Wesbrook Mall, Vancouver, BC V6T 2A3, Canada}

\author{Michael Shamma}
\email{mshamma@triumf.ca}
\affiliation{TRIUMF, 4004 Wesbrook Mall, Vancouver, BC V6T 2A3, Canada}

\begin{abstract}
${}$\vspace{1cm}

New gauge forces can play an important role in the evolution of the early universe. In this work we investigate the cosmological implications of a pure Yang-Mills dark sector that is dominantly populated after primordial inflation. Such a dark sector takes the form of a bath of dark gluons at high temperatures, but confines at lower temperatures to produce a spectrum of dark glueballs. These glueballs then undergo a freezeout process such that the remnant population is nearly completely dominated by the lightest state. To reproduce the observed cosmology, this lightest glueball species must decay to the Standard Model to repopulate and reheat it. At leading order, this can occur through a connector operator of dimension-6. In contrast, other glueballs can be parametrically long-lived or stable, and remain as contributors to dark matter or modify the observed cosmology through their later decays. In this work we study the evolution of such dark sectors in detail. We demonstrate that stable remnant glueballs can produce the measured dark matter abundance. We also derive broad constraints on non-Abelian dark sectors from overproduction of remnant glueballs when they are stable or from their destructive impact when they are able to decay.
\end{abstract}
\maketitle

\section{Introduction}
\label{sec:intro}

Gauge forces underlie the structure of particle physics as described by the Standard Model~(SM). A natural question to ask is whether there exist new gauge forces in addition to the $SU(3)_c\times SU(2)_L\times U(1)_Y$ of the SM (and gravity). Indeed, many proposals for new physics beyond the SM predict new gauge structures~\cite{Pati:1974yy,Georgi:1974sy,Susskind:1978ms,Hill:2002ap,Blumenhagen:2005mu,Langacker:2008yv}. Such new forces can produce a wide range of experimental and astrophysical signatures.

New \emph{dark} gauge forces~\cite{Fayet:2007ua,Pospelov:2008zw,Bjorken:2009mm,Lanfranchi:2020crw,Agrawal:2021dbo,Antel:2023hkf}, that only couple feebly to ordinary matter, are motivated further by the mystery of dark matter~(DM)~\cite{Jungman:1995df,Bertone:2004pz,Lin:2019uvt}. Dark gauge forces can act as mediators between DM and the SM, and play a crucial role in determining the relic DM abundance~\cite{Boehm:2003hm,Pospelov:2007mp,Arkani-Hamed:2008hhe,Feng:2009mn,Chu:2011be,Dvorkin:2019zdi}. A key example is the well-studied dark photon based on a new Abelian gauge group~\cite{Okun:1982xi,Holdom:1985ag}. Dark gauge bosons in other scenarios can bind fundamental constituents into composite DM species~\cite{Nussinov:1985xr,Barr:1990ca,Gudnason:2006yj,Ryttov:2008xe,An:2009vq,Lewis:2011zb,Kribs:2016cew} or even make up the DM itself~\cite{Nelson:2011sf,Arias:2012az,Graham:2015rva}. In this work we study a realization that can do both of these things at once: a theory of dark gluons that confine into dark glueballs that make up DM. 

Dark glueballs arise in the simplest realization of a non-Abelian dark force: a pure Yang-Mills theory based on a non-Abelian Lie group $G_x$. At sufficiently high temperatures $T_x$ the theory consists of dark gluon degrees of freedom. However, at some lower temperature the theory is expected to undergo confinement, with the gluon degrees of freedom binding into glueballs~\cite{Jaffe:1985qp,Mathieu:2008me}. This expectation has been confirmed by lattice studies, which find a thermal confinement phase transition~\cite{Boyd:1996bx,Umeda:2008bd,Asakawa:2013laa,Giusti:2016iqr,Borsanyi:2012ve,Caselle:2018kap,Lucini:2003zr,Lucini:2005vg,Panero:2009tv,Datta:2010sq,Francis:2015lha} and a spectrum of glueballs with masses near the confinement scale as $T_x\to 0$~\cite{Morningstar:1999rf,Chen:2005mg,Meyer:2008tr,Athenodorou:2020ani,Teper:1998kw,Lucini:2012gg}.

Cosmological and DM implications of dark glueballs have been studied in a number of previous works~\cite{Faraggi:2000pv,Juknevich:2009ji,Juknevich:2009gg,Cline:2013zca,Boddy:2014qxa,Yamanaka:2014pva,GarciaGarcia:2015fol,Soni:2016gzf,Forestell:2016qhc,Halverson:2016nfq,Soni:2016yes,Acharya:2017szw,Soni:2017nlm,Mitridate:2017oky,Forestell:2017wov,Kang:2019izi,Asadi:2021pwo,Asadi:2022vkc,Carenza:2022pjd,Carenza:2023eua,Foster:2022ajl,Bishara:2024rtp}. If all the glueballs are stable or very long-lived, they can make up the DM abundance provided the dark sector was populated by inflation (or similar) to a much smaller initial density and temperature than the visible SM sector~\cite{Boddy:2014yra,Soni:2016gzf,Forestell:2016qhc}. The total relic density in this scenario is nearly entirely dominated by the lightest glueball in the spectrum, which generically is expected to have quantum numbers $\jpc=\zpp$~\cite{West:1995ym}. When some of the dark glueballs can decay to the SM, strong bounds can be put on their existence from the impact of the decays on primordial Big-Bang nucleosynthesis~(BBN), the cosmic microwave background~(CMB), and the spectrum of cosmic rays observed today~\cite{Faraggi:2000pv,Cline:2013zca,Boddy:2014yra,Soni:2016gzf,Forestell:2016qhc,Halverson:2016nfq,Forestell:2017wov}. 

An additional signal from the dynamics of a new dark Yang-Mills theory in the early universe is gravitational waves. These can be generated by the confining phase transition of the theory, in which dark gluons bind into dark glueballs, provided the phase transition is strongly first order. Lattice studies of Yang-Mills confinement with $G_x = \sun$ have shown that the transition is first-order for $N\geq 3$ and becomes increasingly strong with larger $N$~\cite{Lucini:2003zr,Lucini:2005vg}. Leveraging these lattice studies for $\sun$ and some other groups, Refs.~\cite{Kubo:2018vdw,Halverson:2020xpg,Huang:2020crf,Kang:2021epo,Morgante:2022zvc} computed the gravitational wave spectrum and demonstrated that the frequency peak could lie within the sensitivity range of proposed future detectors such as DECIGO~\cite{Kawamura:2011zz,Kawamura:2020pcg} and Big Bang Observer~(BBO)~\cite{Corbin:2005ny,Thrane:2013oya} for confinement scales between $1$--$1000\,\gev$, particularly for large values of $N\gg 3$. However, for the signal to be potentially detectable the initial dark gluon density must be the dominant energy component at confinement.

Early dark gluon dominance contrasts with most previous cosmological studies of dark glueballs as DM, which typically assumed a small initial dark gluon energy fraction. Both scenarios can arise from reheating after primordial inflation, depending on whether the inflaton decays preferentially to the dark or visible sector during reheating~\cite{Adshead:2016xxj,Tenkanen:2016jic,Berlin:2016gtr,Adshead:2019uwj,Allahverdi:2020bys}. However, with early dark gluon dominance it is essential that nearly all of the resulting dark glueball density decay to and (re-)reheat the visible sector well before today~\cite{Halverson:2020xpg,Huang:2020crf,Kang:2021epo}. This requires transfer operators that connect between the dark and visible sectors, and implies the dominant lightest $\zpp$ glueball is no longer a viable DM candidate.

In this work we investigate early dark gluon dominance and show that the scenario could still allow dark glueballs to be DM. The only additional structure needed beyond the SM and a dark Yang-Mills sector are transfer operators between them~\cite{Juknevich:2009ji,Juknevich:2009gg}. Among these, the lowest-dimension possibility is the Higgs portal $H^\dagger H\,{\rm tr}(X_{\mu\nu}X^{\mu\nu})$, where $X_{\mu\nu}$ is the dark gluon field strength. This operator allows the lightest and typically dominant $\zpp$ glueball to decay directly to the SM, and also induces transition decays of most of the other glueballs down to this lightest mode. A key exception to this is the lightest $C$-odd glueball, which for $SU(N\geq 3)$ has $\jpc=\opm$. To decay, this state requires a breaking of charge conjugation number in the dark sector $C_x$~\cite{Juknevich:2009ji}. It will therefore be stable if $C_x$ is conserved and a potential DM candidate. Our calculations demonstrate that the $\opm$ glueball can indeed make up the entire DM abundance over a range of masses, even after freezeout and dilution by the decays of the $\zpp$ glueball to the SM. Even when $C_x$ is broken, the leading operator for $\opm$ decay only arises as dimension-8 implying that it is parametrically long-lived relative to the $\zpp$ glueball.

\begin{flushleft}
To be precise, we investigate the following cosmological timeline:
\begin{itemize}
\item Primordial inflation reheats primarily to the dark sector, which self-thermalizes to a dark temperature $T_x$ above the dark confinement temperature $T_c$.
\item The universe expands and cools, some energy is exchanged between the dark and visible sectors through the transfer operators, and eventually the dark sector confines at $T_x=T_c$ to yield a thermal bath of dark glueballs and a visible sector at temperature $T \leq T_c$.
\item The dark glueballs undergo freezeout reactions and decays (when allowed).
\item Reheating of the visible sector occurs when the vast majority of the dark glueballs decay, which is typically the $\zpp$ mode.
\item A density of $\opm$ glueballs is left over as a remnant. These contribute to the DM density if they are stable, or decay at a parametrically later era if $C_x$ is violated and they are unstable.
\end{itemize}
\end{flushleft}

The outline of this paper is as follows. After this introduction we present in Sec.~\ref{sec:gb} the details of the dark sector and its connections to the Standard Model in both the deconfined and confined phases. Next, in Sec.~\ref{sec:rates} we compute the reaction rates needed for the analysis to follow. In Sec.~\ref{sec:before} we investigate the co-evolution of the dark and visible sectors before and up to confinement, while in Sec.~\ref{sec:after} we study glueball freezeout and decay after confinement. In Sec.~\ref{sec:cosmo} we apply our results to study the cosmological implications of the scenario. Finally, Sec.~\ref{sec:conc} is reserved for our conclusions.

\section{Dark Gluons and Glueballs\label{sec:gb}}
 In this section we present essential details of the theory to be studied. We also review lattice results relevant to our cosmological analysis, and estimate glueball interactions and matrix elements.

\subsection{General Setup}

We consider the Standard Model~(SM) augmented by a non-Abelian dark gauge force based on the gauge group $G_x = SU(N\geq 3)$ with the Lagrangian
\beq
\lag = \lag_{\text{SM}} + \lag_x + \lag_{tr} \ .
\eeq
Here, $\lag_\text{SM}$ is the usual SM expression,
while gauge invariance fixes the dark sector portion to be
\beq
\lag_x = -\frac{1}{4}\,X_{\mu\nu}^aX^{a\,\mu\nu} 
+\frac{\alpha_x}{8\pi}\Theta_xX_{\mu\nu}^a\widetilde{X}^{a\,\mu\nu} 
\label{eq:lx}
\eeq
where $X_{\mu\nu}^a$ is the field strength built from the dark gauge field $X_\mu^a$, and $\Theta_x$ is a topological term for the dark gauge theory.

We connect these two sectors through the transfer operators
\beq
-\lag_{tr}
&=& \frac{\kappa_6}{M^2}\,H^{\dagger}H\,X_{\mu\nu}^aX^{a\,\mu\nu} 
+ \frac{\kappa_{8}}{M^4}\,B_{\mu\nu}\Omega^{\mu\nu} + \ldots 
\label{eq:ltr}
\eeq
where $M$ is a large mass scale, $\kappa_{6,8}$ are dimensionless, and $\Omega_{\mu\nu}$ is a linear combination of 
\beq
\Omega_{\mu\nu}  \ \ \supset \ \ {\rm tr}\!\left(X_{\mu\nu} X_{\alpha\beta}X^{\alpha\beta}\right) 
\ \ , \ \
{\rm tr}\!\left(X_\mu^{\alpha} X_{\alpha}^{\beta}X^{\beta\nu}\right)
\ .
\label{eq:nocx}
\eeq
The first operator in Eq.~\eqref{eq:ltr} is the only dimension-6 possibility (up to dualizing one of the field strengths). The second dimension-8 operator is the leading term that breaks the dark-sector charge conjugation invariance $C_x$ respected by all the other operators~\cite{Juknevich:2009ji,Juknevich:2009gg}. There are many other $C_x$ conserving operators at dimension 8, but for our purposes they are subleading relative to the dimension-6 operator~\cite{Juknevich:2009ji,Juknevich:2009gg}.

These transfer operators can arise in several ways. For example, the dimension-6 operator can be generated by a massive scalar with $G_x$ charge and a Higgs portal coupling to the SM Higgs field~\cite{Faraggi:2000pv,Forestell:2016qhc,Forestell:2017wov}, or by new vectorlike fermions with $G_x$ and SM charges and a Yukawa interaction with the Higgs~\cite{Juknevich:2009ji,Juknevich:2009gg}. The dimension-8 operators can be generated by vectorlike fermions charged under both $G_x$ and the SM~\cite{Juknevich:2009ji,Juknevich:2009gg}. Motivated by these scenarios, we fix the dimensionless coefficients in Eq.~\eqref{eq:ltr} to 
\beq
\kappa_6 = \frac{\alpha_x}{4\pi} \ , 
\qquad
\kappa_8 = {\alpha_Y^{1/2}\alpha_x^{3/2}}\,\delta_{C_x} \ , \label{eq:eft-coeffcients}
\eeq
where $\delta_{C_x} = 0,\,1$ depending on whether $C_x$ is conserved or violated. Any further model dependence is absorbed into the dimensionful scale $M$, but this scale should coincide approximately with the mass of new physics generating these operators.
Aside from these coefficients, we do not consider further the ultraviolet~(UV) origin of the connector operators and we assume that the new physics involved does not impact the cosmological signatures (although see Refs.~\cite{Cline:2013zca,Mitridate:2017oky,Asadi:2021pwo,Asadi:2022vkc}).

\subsection{Glueball Data from the Lattice}

Lattice studies of pure $\sun$ gauge theories observe confinement and a spectrum of stable glueballs~\cite{Chen:2005mg,Meyer:2008tr,Lucini:2012gg,Athenodorou:2020ani}. With a minimal action that respects parity and time-reversal, corresponding to a zero theta-term in Eq.~\eqref{eq:lx}, the glueball states in the spectrum can be classified by spin $J$, parity $P$, and charge conjugation $C$. On general grounds the lightest mode is expected to have $\jpc=\zpp$~\cite{West:1995ym}. This has been confirmed by numerous lattice investigations, which find a number of other, heavier, stable glueballs with various $\jpc$ quantum numbers as well~\cite{Morningstar:1999rf,Chen:2005mg,Meyer:2008tr,Lucini:2012gg}. Including the theta-term in Eq.~\eqref{eq:lx} breaks $P$ and $T$ but not $C$, and induces mixing between the would-be $P$-even and $P$-odd states~\cite{Vicari:2008jw}. Simulations at finite temperature also identify a confining phase transition with critical temperature $T_c$ that is first-order for $N\geq 3$~\cite{Lucini:2005vg,Lucini:2003zr,Caselle:2018kap}.

In this work we study $\sunge$ dark Yang-Mills theories with a primary focus on $N=3$ and a small theta-term. All dimensionful quantities in these theories, such as glueball masses, vary in proportion to an overall scale factor. In lattice studies, this scale is typically taken to be the string tension $\sqrt{\sigma}$ or the potential radius $r_0$, which for $N=3$ are related by $r_0\sqrt{\sigma} = 1.160(6)$~\cite{Athenodorou:2020ani}. For our purposes, it is more convenient to use the mass $m_0$ of the lightest $\zpp$ glueball instead. Rescaling lattice results in terms of $m_0 = 3.405(21)\,\sqrt{\sigma}$~\cite{Athenodorou:2020ani}, we find
\beq
\Lambda_{\overline{\rm MS}}= m_0/6.43~\text{\cite{Gockeler:2005rv}} \ ,
\qquad
T_c = m_0/5.30~\text{\cite{Boyd:1996bx,Meyer:2008tr,Francis:2015lha}} \ ,
\qquad
m_1 = 1.78\,m_0~\text{\cite{Athenodorou:2020ani}} \ ,
\label{eq:lattice}
\eeq
where $\Lambda_{\overline{\rm MS}}$ is the confinement scale, $T_c$ is the confinement temperature, and $m_1$ is the mass of the lightest $C$-odd glueball, found to have $\jpc = \opm$. 

While we concentrate on $SU(3)$, we expect qualitatively similar results for $SU(N>3)$ gauge groups. Lattice studies of glueballs for larger $N$ find corresponding states and spectra with a weak mass dependence on $N$ of the form $m \sim A + B/N^2$ for $N$-independent constants $A$ and $B$~\cite{Teper:1998kw,Lucini:2012gg}. The confinement temperature $T_c$ and confinement scale $\Lambda_{\overline{\text{MS}}}$ show a similar dependence~\cite{Lucini:2012gg}. In contrast, many of our findings would not be applicable to $SU(2)$, $SO(2N+1)$, and $Sp(2N)$ gauge groups which do not have $C$-odd glueballs (which require a non-vanishing anomaly coefficient $d^{abc}=tr(t^a\{t^b,t^c\})$)~\cite{Juknevich:2009ji,Juknevich:2009gg}. We also have $SO(4)\simeq SU(2)\times SU(2)$ and $SO(6) \simeq SU(4)$, while for $SO(2N>6)$ the $C$-odd glueballs are expected to be parametrically heavier than the lightest $\zpp$ state~\cite{Juknevich:2009ji,Juknevich:2009gg}.

\subsection{Glueball Self-Interactions and Matrix Elements}

Several glueball matrix elements and interaction vertices are needed to estimate glueball evolution in the early Universe. Whenever possible we obtain them from lattice studies. If they are not available, we estimate them using naive-dimensional analysis~(NDA)~\cite{Manohar:1983md,Cohen:1997rt,Nishio:2012sk} combined with large-$N$ scaling arguments~\cite{tHooft:1973alw,Witten:1979kh}. 

Together, large-$N$ and NDA can be applied to motivate the general form of a low-energy effective Lagrangian that describes the excitations of the theory below confinement. Confinement corresponds to strong coupling with $\alpha_x \to 4\pi/N$ and generates a single dimensionful scale that we identify with the lightest $\zpp$ glueball mass $m_0$~\cite{Witten:1979kh}. The correspondence for the minimal gluon Lagrangian is~\cite{Nishio:2012sk}
\beq
\lag_{x} \ \to \ \lag_{IR} \ = \ \lrf{\zeta N}{4\pi}^2\!m_0^4\,\mathcal{F}(\hat{\phi}/m_0,\del/m_0) \ , \nnmb
\eeq
where $\hat{\phi}$ is a generic (unnormalized) glueball field, $\mathcal{F}$ is an unspecified function that respects the underlying symmetries of the theory, and $\zeta$ is a dimensionless factor of order unity that we use to parameterize uncertainties within this approach. Expanding in powers of fields and rescaling to obtain canonically normalized glueball fields $\phi$, we obtain
\beq
\lag_{IR} \ \to \ \frac{1}{2}(\del\phi)^2 - \sum_{n=1}^{\infty}\frac{a_n}{n!}\!\lrf{4\pi}{\zeta N}^{\!n-2}\!\!m_0^{4-n}\phi^n \ ,
\label{eq:lnda1} 
\eeq
for coefficients $a_n$ of order unity. This effective Lagrangian is expected to include kinetic and mass terms for all glueballs as well as all interactions conserving $C_x$ (but possibly violating $P$ and $T$ for non-zero $\Theta_x$). We will use it below to estimate reaction rates among glueballs.

With large-$N$ and NDA we can also estimate glueball matrix elements involving the connector operators of Eq.~\eqref{eq:ltr}. The key transition matrix element of interest for the direct decay of the $\zpp$ glueball $\phi_0$ to the SM through the Higgs portal operator is
\beq
\alpha_x F_0 \ \equiv \ \frac{\alpha_x}{2}\left<0\right|X_{\mu\nu}^aX^{a\,\mu\nu}\left|\zpp\right> \ \sim \ m_0^3 \ ,
\eeq
where the last estimate comes from NDA and is independent of $N$.
This matrix element has also been computed on the lattice with the result~\cite{Chen:2005mg,Meyer:2008tr} (rescaled with the $\zpp$ mass of Ref.~\cite{Athenodorou:2020ani})
\beq
\alpha_xF_0 = 0.215\,m_0^3 \ ,
\eeq
The NDA estimate above is reasonably consistent with this lattice determination. We use the lattice result in the calculations to follow.

If dark sector charge conjugation $C_x$ is violated, there will also be decays of the lightest $C$-odd $\opm$ glueball $\opm \to \zpp+\gamma/Z$ through the dimension-8 operator of Eq.~\eqref{eq:ltr}. The relevant matrix element is 
\beq
\left<\zpp(p')\right|\Omega_{\mu\nu}\left|\opm(p,\lambda)\right>
\ \equiv \ \epsilon_{\mu\nu\alpha\beta}\,p^\alpha\varepsilon^{\beta}(p,\lambda)\;M_{10}  \ ,
\eeq
where $\Omega_{\mu\nu}$ is the 3-glueball operator defined in Eq.~\eqref{eq:nocx}. The matrix element factor $M_{10}$ has not been computed on the lattice so we estimate its value with NDA:
\beq
\alpha_x^{3/2}M_{10} \ \sim \ \frac{\sqrt{4\pi}}{N}\,m_0^3 \ .
\eeq

\section{Reaction Rates Between and Within Sectors
\label{sec:rates}}

The connector operators of Eq.~\eqref{eq:ltr} induce scattering and decay processes that transfer particles and energy between the dark and visible sectors before and after dark confinement. In the confined phase, glueball interactions described by Eq.~\eqref{eq:lnda1} also produce number-changing reactions that control the resulting relic glueball densities. We compute the dominant reaction rates for these processes in this section.

\subsection{Transfer Rates in the Deconfined Phase}

At higher dark temperatures, $T_x \gg m_0$, the dark sector is described well by a nearly free dark gluon gas. In this deconfined phase it is sufficient to track the dark and visible sector energy densities, $\rho_x$ and $\rho$. We expect and assume that both sectors are self-thermalized with temperatures $T_x$ and $T$, respectively. 

Transfer reactions induced by the operators of Eq.~\eqref{eq:ltr} contribute to the evolution of $\rho_x$ and $\rho$ through an integrated energy collision term $\mathcal{C}_E^{tr}$ according to
\beq
\frac{d\rho_x}{dt} \ \supset \ -\mathcal{C}_E^{tr} \ ,\qquad
\frac{d\rho}{dt} \ \supset \ +\mathcal{C}_E^{tr} \ .
\label{eq:etrans1}
\eeq
For $T_x > T$ above the weak sale, the dominant reaction contributing to $\mathcal{C}_E$ is $X^\mu(p_1)+X^\nu(p_2) \to H(p_3)+H^{\dagger}(p_4)$. The energy transfer collision term for this reaction is
\beq
\mathcal{C}_E^{tr} &\simeq& \frac{1}{2}\!\int\!d\Pi_1\ldots\int\!d\Pi_4\;(E_1+E_2)\,(2\pi)^4\delta^{(4)}(\sum_ip_i)\,|\mathcal{M}|^2\,(f_1f_2-f_3f_4) 
\label{eq:ce}\\
&\simeq& \frac{0.31}{(4\pi)^2}\,(N^2-1)\!\left[\alpha_x^2(T_x)\frac{T_x^9}{M^4} - \alpha_x^2(T)\frac{T^9}{M^4}\right] \ .
\nnmb
\eeq
where the first term is a symmetry factor for indistinguishable initial states, $(E_1+E_2) = (E_3+E_4)$ is the energy transferred out of the dark sector, the matrix element $|\mathcal{M}|^2$ is summed over all initial and final states, $d\Pi_i = d^3p_i/2E_i(2\pi)^3$,
and we have neglected quantum correction factors that we expect to be small. In the second line we have evaluated the thermal averages in the Maxwell-Boltzmann approximation for the explicit dimension-6 transfer operator of Eq.~\eqref{eq:ltr}. Note that the result depends on the running value of $\alpha_x$, and for this we use the known two-loop expression with running scale $\Lambda_x$ given in Eq.~\eqref{eq:lattice} but truncated at $\alpha_x \leq 4\pi/N$, as suggested by several strong-coupling estimates~\cite{Deur:2016tte}.

\subsection{Transfer Rates in the Confined Phase}

In the confined phase the dark sector consists of a spectrum of glueballs. Here, it is important to track the the number densities of (some) individual glueballs in addition to total energy densities of the dark visible sectors. For the number density of the $i$-th glueball species, we have
\beq
\frac{dn_i}{dt} \ \supset \  - \mathcal{C}_{n_i}^{x}  -  \mathcal{C}_{n_i}^{tr} \ ,
\eeq
where $\mathcal{C}_{n_i}^{x}$ describes glueball depletion by reactions within the dark sector and $\mathcal{C}_{n_i}^{tr}$ corresponds to number-changing transfer processes. Energy transfer impacts the energy densities as in Eq.~\eqref{eq:etrans1}, although with a different expression for the energy collision term $\mathcal{C}_E^{tr}$.  In this subsection we compute the dominant contributions to $\mathcal{C}_E^{tr}$ and $\mathcal{C}_{n_i}^{tr}$, leaving $\mathcal{C}_{n_i}^{x}$ to the next.

The dominant reactions in the confined phase for both number and energy transfer are glueball decays. The rate of number change from decays is well known in the Maxwell-Boltzmann approximation~\cite{Kolb:1979qa}. Generalizing slightly to the case of two sectors, each in equilibrium with itself but not necessarily with the other, we find
\beq
\mathcal{C}_{n_i}^{tr} = {\Gamma}_i\left[\frac{K_1(x_i)}{K_2(x_i)}\,n_i(x_i) - \frac{K_1(\tilde{x}_i)}{K_2(\tilde{x}_i)}\,\bar{n}_i(\tilde{x}_i)\right] \ ,
\label{eq:cntr}
\eeq
where $\Gamma_i$ is the total decay width of the $i$-th glueball to SM final states, $K_\alpha$ is the modified Bessel function of order $\alpha$, $x_i= m_i/T_x$, $\tilde{x}_i=m_i/T$, and $\bar{n}_i$ is the equilibrium number density. The ratios $K_1(x_i)/K_2(x_i) = \langle m_i/E_i\rangle= \langle 1/\gamma\rangle$ correspond to thermal averages of time dilation factors relative to the plasma rest frame. Generalizing further to describe energy transfer is straightforward. We find
\beq
\mathcal{C}_{E}^{tr} = m_i{\Gamma}_i\bigg[n_i(x_i) - \bar{n}_i(\tilde{x}_i)\bigg] \ .
\label{eq:cetr}
\eeq
Interestingly, the factors of the energy transferred cancel precisely with the energy factor in the time dilation $1/\gamma = m_i/E_i$.

Among the many glueball species, the two specific states that we will be the most interested in are the lightest $C_x$-even $\zpp$ and the lightest $C_x$-odd $\opm$ modes. The $\zpp$ glueball decays directly to the SM through the dimension-6 Higgs portal operator of Eq.~\eqref{eq:ltr}. For $m_0 \gtrsim 2\,m_h$, its decay width scales like $\Gamma_0 \sim m_0^5/M^4$. This operator (or its $PT$ conjugate) also mediates the decays of all other glueballs down to the $\zpp$ with one exception. Specifically, the $\opm$ glueball is stable if dark sector charge conjugation number $C_x$ is preserved. When $C_x$ is broken, the dimension-8 operator of Eq.~\eqref{eq:ltr} is allowed and mediates $\opm \to \zpp + \gamma/Z$ with decay width $\Gamma_1 \sim m_0^9/M^8$, parametrically slower than the decay rate of the $\zpp$ glueball.

Let us now examine these rates more carefully. In the absence of electroweak symmetry breaking, there is the direct decay $\zpp \to H+H^{\dagger}$ with rate
\beq
\Gamma_0
= \frac{1}{32\pi^3}\!\lrf{\alpha_xF_0}{m_0^3}^{\!2}\!\frac{m_0^5}{M^4} \ .
\eeq
In the electroweak broken phase with $m_0< 2m_h$, the $\zpp$ decays via an off-shell Higgs boson, $\zpp \to h^* \to SM$. Writing $H\to(0,(v+h)/\sqrt{2})^t$, the decay rate is~\cite{Juknevich:2009gg} 
\beq
\Gamma_0(\zpp \to h^* \to \text{SM}) = 
\frac{1}{4\pi^2}\frac{(\alpha_x F_0)^2 v^2}{M^4[(m_0^2-m_h^2)^2 + (m_h\Gamma_h)^2]}\,\Gamma_h(m_h\to m_0)
\eeq
where $m_h=125\,\gev$ and $\Gamma_h=4.1\,\mev$ are the mass and decay width of the Higgs boson, and $\Gamma_h(m_h \to m_0)$ stands for the decay width of the off-shell Higgs boson with energy $m_0$. When $m_0 > 2m_h$, a further decay channel $\zpp\to hh$ opens up with partial width ~\cite{Juknevich:2009gg}
\beq
\Gamma_0(\zpp \to hh) = \frac{1}{128\pi^3}\lrf{\alpha_x^2 F_0^2}{m_0^3}^{\!2}\!\frac{m_0^5}{M^4}\!\left(1 + \frac{3m_h^2}{|(m_0^2-m_h^2)+i m_h\Gamma_h|} \right)^{\!2}\!\sqrt{1 - \frac{4m_h^2}{m_0^2}}
\eeq
We also note that for $m_0 \gg 2 m_h$, the total decay width of the $\zpp$ approaches the width in the electroweak unbroken phase and the partial widths asymptote to
\beq
\Gamma_{WW} : \Gamma_{ZZ} : \Gamma_{hh} \sim 2: 1: 1
\eeq
Both features are in accordance with the Goldstone equivalence principle~\cite{Cornwall:1974km,Vayonakis:1976vz,Lee:1977yc,Chanowitz:1985hj}.

Turning to the lightest $C_x$-odd state, the $\opm$ glueball, if $C_x$ is broken it can decay via the dimension-8 operator in Eq.~\eqref{eq:ltr} through the channels $\opm \to \zpp + \gamma$ and $\opm \to \zpp + Z$. The partial widths are~\cite{Juknevich:2009gg,Juknevich:2009ji} 
\beq
\Gamma(\opm \to \zpp + \gamma) &=& \frac{\alpha}{24\pi}\!\lrf{\alpha_x^{3/2} M_{10}}{m_0^3}^{\!2}
\!\frac{m_1^3m_0^6}{M^8}
\left( 1 - \frac{m_0^2}{m_1^2} \right)^{\!3} \\
\Gamma(\opm \to \zpp + Z) &=& \frac{\alpha}{24\pi}\,t_W^2
\lrf{\alpha_x^{3/2} M_{10}}{m_0^3}^{\!2}\frac{m_1^3m_0^6}{M^8}
\left[ \left(1 + \frac{m_0^2}{m_1^2} - \frac{m_{Z}^2}{m_1^2}\right)^2-4\,\frac{m_0^2}{m_1^2}\right]^{3/2}
\eeq
where $\alpha$ is the electromagnetic coupling and $t_W$ is the tangent of the Weinberg angle.

\subsection{Reaction Rates within the Confined Dark Sector}

When the lifetimes for decays to the SM are long, glueballs undergo a freezeout process after confinement. To compute the evolution of the glueball sector, several glueball annihilation rates are needed. These can be estimated with the glueball effective Lagrangian of Eq.~\eqref{eq:lnda1}. 

To track the impact of dark glueballs on the evolution of the cosmos, we will focus on the total glueball density together with the specific density of the $\opm$ state. The total glueball density is simply the sum of the densities of all the glueballs that are stable in the absence of connector operators,
\beq
n_x = \sum_in_i \ .
\eeq
Tracking $n_x$ is convenient because it is insensitive to the many 2-to-2 annihilation and co-annihilation reactions amongst the dark glueballs. Instead, the evolution of $n_x$ (in the absence of decays) is dominated by 3-to-2 reactions involving the $\zpp$ state. The corresponding collision term is~\cite{Carlson:1992fn,Hochberg:2014dra}
\beq
\mathcal{C}_{n_x}^{x} \ \simeq \ \langle\sigma_{32}v^2\rangle\,n_0^2\left[n_0-\bar{n}_0(x)\right] \ ,
\label{eq:cnx}
\eeq
where $\langle\sigma_{32}v^2\rangle$ is the thermal average of the $3\{\zpp\}\to 2\{\zpp\}$ cross section. Using the effective Lagrangian of Eq.~\eqref{eq:lnda1}, we estimate
\beq
\langle\sigma_{32} v^2\rangle \ \simeq \ \frac{1}{(4\pi)^3}\lrf{4\pi}{\zeta N}^{\!6}\frac{1}{m_0^5}
\label{eq:sig32}
\eeq
To avoid large power enhancements for small $N$, we set the parameter $\zeta = 4$ in the analysis to follow. Recall that its specific value represents the theoretical uncertainty in the large-$N$/NDA estimate of glueball couplings.

For the $\opm$ glueball, the dominant annihilation reaction at late times is $\opm+\opm \to \zpp+\zpp$ annihilation. In contrast to the $CP$-even glueballs, the $\opm$ is unable to coannihilate with the $\zpp$ glueball due to $C_x$ conservation in the dark sector. The leading collision term for the evolution of the $\opm$ number density is thus
\beq
\mathcal{C}_1^{x} \ \simeq \ \langle\sigma_{22}v\rangle\!\left[n_1^2-\lrf{n_0}{\bar{n}_0}^2\!\bar{n}_1^2 \right] \ .
\label{eq:c1x}
\eeq
In this expression we see an enhancement of the reverse reaction when the $\zpp$ glueball is overabundant. To estimate the thermally averaged cross section, we again apply Eq.~\eqref{eq:lnda1} with the same not-so-large $N$ modification as described above to obtain
\beq
\langle\sigma_{22} v\rangle \ \simeq \ \frac{1}{(4\pi)}\lrf{4\pi}{\zeta N}^{\!4}\frac{1}{m_0^2}  \ .
\label{eq:sig22}
\eeq
As above, we set $\zeta = 4$ in the analysis to follow. Note as well that this cross section is representative of the elastic scattering cross section among dark glueballs.


\section{Evolution Before and Through the Confining Transition
\label{sec:before}}

Recall that our assumption about the very early universe is that it was at some point dominated by dark gluons with temperature $T_x$ above the dark confinement temperature $T_c$, with a subleading population of visible SM states. In this phase, and assuming self-thermalization within each sector, the two sectors can be fully characterized by their energy densities $\rho_x$ and $\rho$, or equivalently by their temperatures $T_x$ and $T$ with $T_x \geq T$. The energy densities evolve according to
\beq
\frac{d\rho_x}{dt} &=& -4H\rho_x\bigg(1 - \frac{\Delta_x}{4\rho_x}\bigg) - \mathcal{C}_E
\label{eq:drxdt}\\
\frac{d\rho}{dt} &=& -4H\rho\bigg(1 - \frac{\Delta}{4\rho}\bigg) + \mathcal{C}_E
\label{eq:drvdt}
\eeq
where $H = \sqrt{(\rho_x+\rho)/3\mpl^2}$ is the Hubble rate with $\mpl$ the reduced Planck mass, $\Delta_i = \rho_i-3p_i$ is the deviation from conformality, and $\mathcal{C}_E$ is the energy transfer rate derived from the collision term in the relevant Boltzmann equations and computed in Sec.~\ref{sec:rates}. 

To solve Eqs.~(\ref{eq:drxdt},\ref{eq:drvdt}), and evaluate $\mathcal{C}_E$, we need expressions for $\Delta_{(x)}$ and $T_{(x)}$ in terms of $\rho_x$ and $\rho$. These relations are known for the visible sector~\cite{Saikawa:2018rcs,Saikawa:2020swg}, but are more complicated for the dark sector as $T_x$ approaches $T_c$. When $T_x \gg T_c$ the dark sector is described well by a gas of free dark gluons with $\Delta_x \simeq 0$ and $\rho_x \simeq \pi^2g_{*\,x}T_x^4/30$ with $g_{*\,x} = 2(N^2-1)$ for $G_x=\sun$. However, as $T_x$ approaches $T_c$ there are important deviations from the free gluon gas picture. In this section we turn to lattice studies to characterize the equation of state and temperature of the dark sector approaching confinement. We then apply these results to compute the evolution of the dark and visible energy densities up to confinement.

\subsection{Dynamics of the Dark Confining Transition}

Thermodynamics of the confining phase transition in $\sun$ Yang-Mills theories have been studied extensively on the lattice~\cite{Boyd:1996bx,Umeda:2008bd,Asakawa:2013laa,Giusti:2016iqr,Borsanyi:2012ve,Caselle:2018kap,Lucini:2003zr,Lucini:2005vg,Panero:2009tv,Datta:2010sq,Francis:2015lha}. For our analysis we take guidance primarily from the $SU(3)$ studies in Ref.~\cite{Borsanyi:2012ve}~(lattice~1) and Ref.~\cite{Caselle:2018kap}~(lattice~2). These works compute the pressure $p_x$ and energy density $\rho_x$ as functions of $T_x/T_c$ at temperatures above and below the confinement temperature. Recall that we also use the determinations of the confinement temperature of Refs.~\cite{Boyd:1996bx,Meyer:2008tr,Francis:2015lha} and express it in terms of the lightest $\suthree$ glueball mass, yielding $T_c = m_0/5.3$.

\begin{figure}[ttt]
  \begin{center}
    \includegraphics[width = 0.47\textwidth]{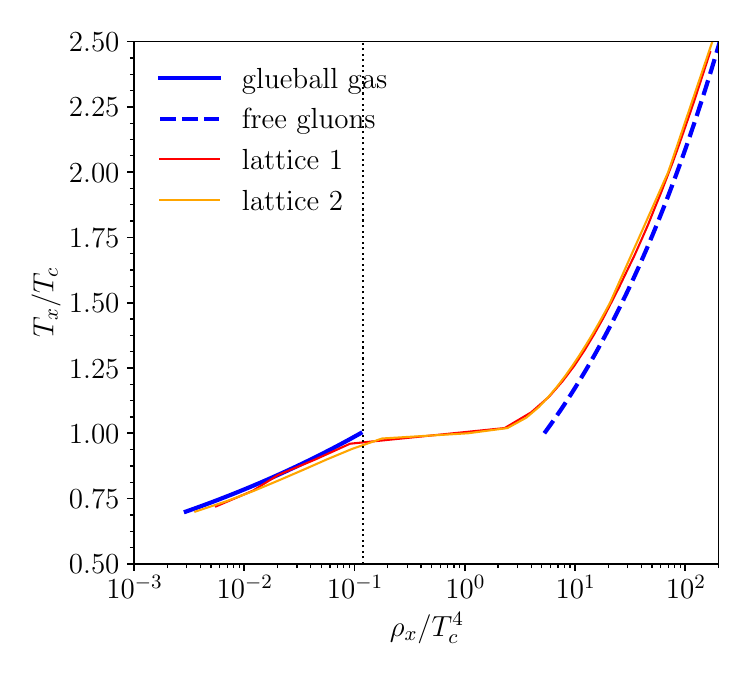}
    \includegraphics[width = 0.47\textwidth]{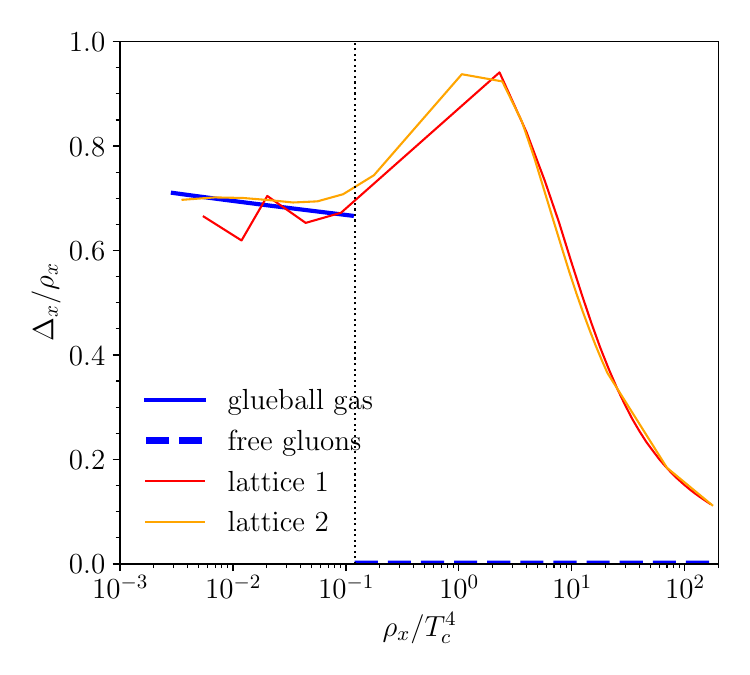}
    \vspace{-0.5cm}
  \end{center}
  \caption{Evolution of $T_x/T_c$~(left) and $\Delta_x/\rho_x$~(right) with $\rho_x/T_c^4$  obtained from the lattice studies of Ref.~\cite{Borsanyi:2012ve}~(lattice~1) and Ref.~\cite{Caselle:2018kap}~(lattice~2), as well as the predictions for a free glueball gas below $T_c$ and free gluons above $T_c$. The vertical dotted line shows our estimate of $\rho_x/T_c^4$ at confinement based on the free glueball gas picture using the spectrum of Ref.~\cite{Athenodorou:2020ani}.
  \label{fig:gbtd}}
\end{figure}

In Fig.~\ref{fig:gbtd} we plot $T_x/T_c$~(left) and $\Delta_x/\rho_x$~(right) as functions of $\rho_x/T_c^4$ near confinement extracted from the Refs.~\cite{Caselle:2018kap,Borsanyi:2012ve}. Relative to a free gluon gas, shown in both panels for $T_x > T_c$ with a dashed blue line, the dark temperature $T_x$ falls to near $T_c$ and stays close to constant as $\rho_x$ decreases further. This is to be expected on average due to the release of latent heat as the first order phase transition proceeds. The conformality parameter $\Delta_x = \rho_x-3p_x$ in the right panel shows a very strong deviation from the free gluon value of zero as $T_x$ approaches $T_c$. We also see that the lattice results trend towards the free gluon limit at temperatures well above $T_c$.

At temperatures below $T_x < T_c$, the lattice results in Fig.~\ref{fig:gbtd} agree remarkably well with the predictions for a free glueball gas computed using the full glueball spectrum of Ref.~\cite{Athenodorou:2020ani}, shown by the blue solid lines in both panels. This was demonstrated previously in Ref.~\cite{Trotti:2022knd}, and motivates us to use the glueball gas description immediately after confinement. 
Based on the glueball gas picture, we also identify $\rho_x/T_c^4 =0.12$ as the energy density at confinement. This value is markedly different from the free gluon gas prediction of $\rho_x/T_c^4 = 8\pi^2/15 \simeq 5.26$.

These lattice results allow us to evaluate the coupled evolution equations of Eqs.~(\ref{eq:drxdt},\ref{eq:drvdt}). We use fits to $\Delta_x/\rho_x$ and $T_x/T_c$ as functions of $\rho_x$ to evaluate Eq.~\eqref{eq:drxdt} at dark temperatures above the confinement scale. As $T_x$ falls below $T_c$, we match to the free glueball picture and identify $\rho_x/T_c^4 = 0.12$ at confinement. For the visible sector, we use the results of Ref.~\cite{Saikawa:2018rcs,Saikawa:2020swg} to extract $T$ and $\Delta$ from $\rho$.

\subsection{Energy Transfer Up to the Confining Transition}

Equipped with these lattice results, we turn now to evaluating Eqs.~(\ref{eq:drxdt},\ref{eq:drvdt}) for $T_x > T_c$. Recall our assumption about the initial conditions that the dark sector is reheated preferentially by primordial inflation. In the absence of transfer reactions, $\mathcal{C}_E \to 0$, this implies that $\rho_x \gg \rho$ and $T_x \gg T$ early on. The energy densities in both sectors would then have mostly redshifted together until $T_x$ approaches $T_c$ and $\Delta_x/\rho_x$ grows, slowing the dilution and increasing the density of the dark sector relative to the visible even further. 

This simple picture can be modified by transfer reactions in a way that is sensitive to the details of primordial reheating. The energy transfer rate $\mathcal{C}_E$ in Eq.~\eqref{eq:ce} increases very rapidly with dark temperature $T_x$, and thus energy transfer is dominated by temperatures near reheating after inflation, $T_x \simeq T_{x,RH}$. If $T_{x,RH}$ was high enough, such that $\mathcal{C}_E \gtrsim H$ at this temperature, transfer reactions would have equilibrated the dark and visible sectors with $T=T_x$. In contrast, the strong UV dependence of $\mathcal{C}_E$ means that the two sectors could have avoided thermalization with each other for lower reheating temperatures, even for the same model parameters $m_0$ and $M$. 

These considerations motivate two distinct sets of initial conditions that bound the full set of possibilities assuming preferential reheating to the dark sector. We begin our numerical evolution at an initial temperature $T_{x,i}=5\,T_c \simeq m_x$, which is near the confinement temperature but high enough that the free gluon picture is a good approximation for the dark sector. The dark sector energy density at $T_{x,i}$ is then set according to our fit to the lattice data. For the visible sector energy density, we consider two cases: i) 
$\rho=(g_*/g_{*x})\rho_x$ corresponding to early thermalization following reheating; ii) $\rho=(0.01)\rho_x$ corresponding to non-thermalization of the two sectors and a strongly sub-dominant visible sector density.

\begin{figure}[!ttt]
  \begin{center}
    \includegraphics[width = 0.47\textwidth]{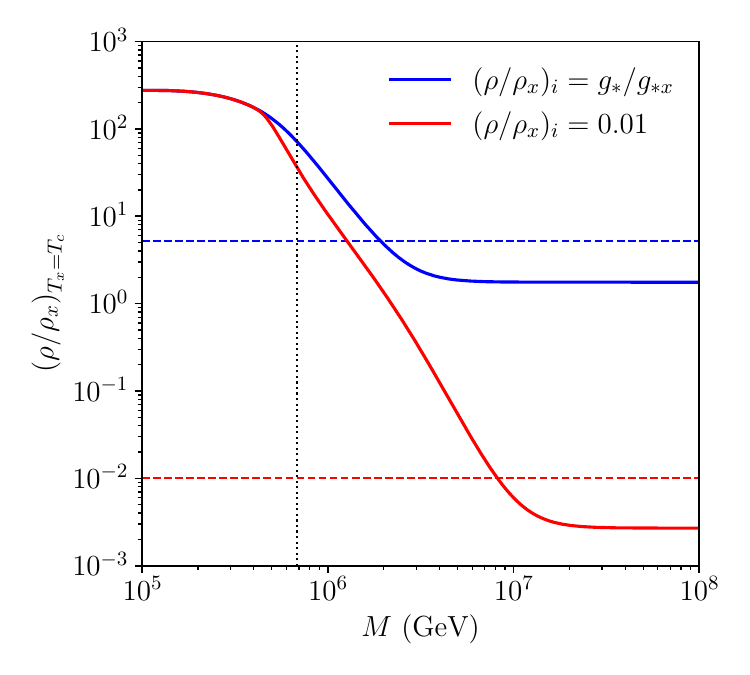}
    \vspace{-0.5cm}
  \end{center}
  \caption{Ratio of SM to dark gluon energy densities $\rho/\rho_x$ at $T_x=T_c$ for $m_0=10^3\,\gev$ assuming an initial dark temperature of $T_{x,i}= 5\,T_c$ and initial ratios $(\rho/\rho_x)_i=g_*/g_{*x}$~(solid blue line) or $(\rho/\rho_x)_i=0.01$~(solid red line). The dashed blue and red lines show the energy ratios in the absence of phase transition dynamics.  The vertical dotted line indicates where $[\mathcal{C}_E/H\rho_x]_{T_{x,i}} = 1$.
  }
  \label{fig:glutrans}
\end{figure}

Our numerical results for the ratio $\rho/\rho_x$ at $T=T_c$ after evolving from $T_{x,i}=5 T_c$ for the two initial conditions discussed above are shown in Fig.~\ref{fig:glutrans}. This ratio is shown as a function of the transfer operator mass scale $M$ defined in Eq.~\eqref{eq:ltr} for a lightest glueball mass of $m_0 = 10^3\,\gev$. However, if the SM can be treated as fully relativistic, there is only one physical scale and these curves also apply to larger $m_0$ values with the rescaling $M\to M\times (m_0/10^3\,\gev)^{3/4}$. Also shown on the plot with horizontal dashed lines are the energy ratios obtained without considering the dynamics of the phase transition. The vertical dotted lines indicate where $[\mathcal{C}_E/H\rho_x]_{T_{x,i}} = 1$ where $\mathcal{C}_E$ only includes the forward dark to visible reaction. This figure demonstrates the effects of energy transfer as well the impact of deviations from the gluon gas approximation approaching $T_c$.

Smaller values of $M$ in Fig.~\ref{fig:glutrans} correspond to faster energy transfer. When $[\mathcal{C}_E/H\rho_x]_{T_{x,i}} \gtrsim 1$ the two sectors thermalize nearly immediately at $T_{x,i}$. This is seen in Fig.~\ref{fig:glutrans} to the left of the vertical dotted line, where the curves for the two limiting initial conditions come together as a result of this fast thermalization. The ratio of $(\rho/\rho_x)_{T_c}$ in this smaller-$M$ region is also considerably larger than $g_*/g_{*x}$. Here, the dark and visible sectors remain equilibrated with $T\simeq T_x$ as $T_x$ approaches $T_c$ while $\rho_x$ continues to fall (as seen in the right panel of Fig.~\ref{fig:gbtd}). For very small $M$, such that thermalization persists all the way to confinement, the ratio asymptotes to $\rho(T=T_c)/\rho_x(T_c)$.

For larger values of $M$, to the right of the vertical dotted line in Fig.~\ref{fig:glutrans}, thermalization is incomplete at $T_{x,i}$ and the ratio of energy densities at $T_x=T_c$ depends on the initial conditions. As $M$ grows even larger, there is no significant thermalization at all between $T_{x,i}$ and $T_c$, and both curves approach constant values. Both are smaller than the initial ratios at $T_{x,i}$ due to the reduced dilution of the dark sector resulting for $T_x$ near $T_c$.

\subsection{Energy Transfer After Confinement}

Energy transfer between the dark and visible sectors after confinement is dominated by dark glueball decays and inverse decays to SM states, as described in Sec.~\ref{sec:rates}. The relevant decay and other interaction rates depend crucially on the lightest glueball mass, and in computing these rates we use the zero temperature value of this mass. However, very near the confinement temperature $T_c$ the masses of glueballs may differ significantly from their values at $T=0$, as suggested by flux tube models of glueballs~\cite{Isgur:1984bm,Caselle:2013qpa}. We argue here that despite this possibility, our use of fixed zero-temperature masses is expected to be a good approximation for computing dark glueball energy transfer and freezeout.

Interpretations of lattice studies of glueball masses at finite temperature do not appear to be conclusive. The studies of $SU(3)$ Yang-Mills in Refs.~\cite{Ishii:2002ww,Meng:2009hh} find that the low-lying glueball masses remain relatively constant as $T$ increases from zero to near $T_c$. In contrast, in Ref.~\cite{Caselle:2013qpa} it is argued that a new mass scale arises near $T_c$ and that when this is accounted for the finite-temperature glueball masses should scale proportionally to the square root of the effective string tension, $m(T)\propto \sqrt{\sigma(T)}$, as occurs in flux tube models~\cite{Isgur:1984bm}. Since this tension is found to decrease significantly approaching $T_c$ (from below)~\cite{Kaczmarek:1999mm,Cardoso:2011hh}, this suggests a significant corresponding reduction in the glueball masses just below confinement.

In our study we simply fix the glueball masses at their $T=0$ values for all $T_x < T_c$. This procedure is supported by the very good agreement between lattice determinations of thermodynamic variables~\cite{Borsanyi:2012ve,Caselle:2018kap} and the predictions of the free glueball gas model all the way up to $T_c$ using the spectrum of zero-temperature glueballs found in Refs.~\cite{Athenodorou:2020ani}, as demonstrated in Ref.~\cite{Trotti:2022knd} and shown in Fig.~\ref{fig:gbtd}. Moreover, even if the glueball masses decrease for $T_x$ near $T_c$ in proportion to the root of the string tension, they would be expected return quickly as $T_x$ falls to nearly their zero temperature values (\emph{e.g.} $m(T)/m(0) \gtrsim 0.8$  for $T_x/T_c \lesssim 0.8$)~\cite{Cardoso:2011hh}. Since the cosmological evolution of glueballs typically occurs over a much larger temperature range, such a change in mass would not significantly affect our results.

\section{Dark Glueball Freezeout
\label{sec:after}}

Confinement binds dark gluons into dark glueball bound states. These glueballs are depleted by dilution, as well as by annihilation and decay reactions. If decays can be neglected, the total number of glueballs per comoving volume can only be reduced by $3\to 2$ and higher reactions~\cite{Carlson:1992fn}. While these occur, heavier glueballs can annihilate to the lightest $\zpp$ mode through parametrically faster $2\to 2$ reactions~\cite{Pappadopulo:2016pkp,Forestell:2016qhc,Farina:2016llk,Forestell:2017wov}. The net result is a freezeout process where the glueballs cool slowly and become dominated by the lightest $\zpp$ state, with exponentially smaller populations of the heavier glueballs. In this section we compute these freezeout dynamics in detail while taking into account the impact of decays. Our focus is on the total glueball density as well as the specific density of the parametrically long-lived or stable $\opm$ glueball.

\subsection{Evolution of the Total Glueball Density}

Tracking the total glueball density provides a convenient way to compute the impact of the dark sector on the evolution of the visible sector. As pointed out in Sec.~\ref{sec:rates}, the total glueball density is also insensitive to the many $2\to 2$ reactions among individual glueballs, but depends instead on $3\to 2$ and higher reactions. This treatment reproduces the features of previous analyses that focused on the $\zpp$ density exclusively while also taking into account the impacts of heavier states.

Throughout the freezeout of the total glueball density we expect to have both kinetic equilibrium and the equilibrium ratios of glueball species $n_i/n_0$ since the $2\to 2$ processes that maintain these are parametrically faster than the $3\to 2$ reactions that control the total density. The latter implies ${n_i}/{\bar{n}_i} = n_0/\bar{n}_0$, where the bar indicates full thermal equilibrium at temperature $T_x$ with zero chemical potential. Using these relations, the dark sector number density, pressure, and energy density are given by
\beq
n_x &=& \sum_in_i = n_0 \times f_n(x) \ ,
\label{eq:nx}\\
p_x &=& \frac{1}{x}\,m_0\,n_x \ ,
\label{eq:px}\\
\rho_x &=& \sum_i\rho_i = m_0\,n_x \times f_\rho(x) \ ,
\label{eq:rhox}
\eeq
with $x= m_0/T_x$, the sums run over all glueball species,
and (in the Maxwell-Boltzmann approximation)
\beq
f_n(x) &=& \sum_ig_i\xi_i^2K_2(\xi_ix)\Big/K_2(x) \ ,
\label{eq:fn}\\
f_\rho(x) &=& 
\sum_ig_i\xi_i^3K_3(\xi_ix)\Big/\sum_jg_j\xi_j^2K_2(\xi_jx) - \frac{1}{x} \ ,
\label{eq:fr}
\eeq
where $\xi_i \equiv m_i/m_0$ is the ratio of masses and $g_i$ is the number of spin degrees of freedom. For $x\gg 1$ we have $f_n \simeq f_\rho \simeq 1$.

Using the form of Eq.~\eqref{eq:rhox} in the evolution equation for the dark sector energy density of Eq.~\eqref{eq:drxdt}, we obtain coupled evolution equations for the dark temperature $x$ and the total glueball density $n_x$:
\beq
g_\rho\frac{dx}{dt} &=& \frac{3p_x}{m_0\,n_x}H - f_\rho\,\frac{\mathcal{C}_n^x}{n_x} - \lrf{f_\rho\mathcal{C}_n^{tr}-\mathcal{C}_E^{tr}/m_0}{n_x} \ ,
\label{eq:dxdt}\\
\frac{dn_x}{dt} &=& -3Hn_x - \mathcal{C}_n^x-\mathcal{C}_n^{tr} \ .
\label{eq:dnxdt}
\eeq
Expressions for $\mathcal{C}_n^x$, $\mathcal{C}_n^{tr}$, and $\mathcal{C}_E^{tr}$ are given in Eqs.~(\ref{eq:ce},\ref{eq:cntr},\ref{eq:cnx}) while
\beq
g_\rho(x) &\equiv & -f^\prime_\rho(x)
\label{eq:grho}\\
&=& \frac{1}{x}\,f_\rho(x) 
+ \sum_ig_i\xi_i^4K_2(\xi_i x)\Big{/}\sum_jg_j\xi_j^2K_2(\xi_jx)
\nnmb\\
&&~~ -\sum_ig_i\xi_i^3K_3(\xi_ix)\,\sum_jg_i\xi_j^3K_1(\xi_ix)\Big{/}\Big[\sum_kg_k\xi_k^2K_2(\xi_kx)\Big]^2
\nnmb
\eeq
For $x\gg 1$ we find $g_\rho(x) \simeq 3/2x^2$. The visible energy density evolves by Eq.~\eqref{eq:drvdt}, which combined with Eqs.~(\ref{eq:dxdt},\ref{eq:dnxdt}) fully characterize the energy evolution of both sectors.

\begin{figure}[!ttt]
  \begin{center}
    \includegraphics[width = 0.47\textwidth]{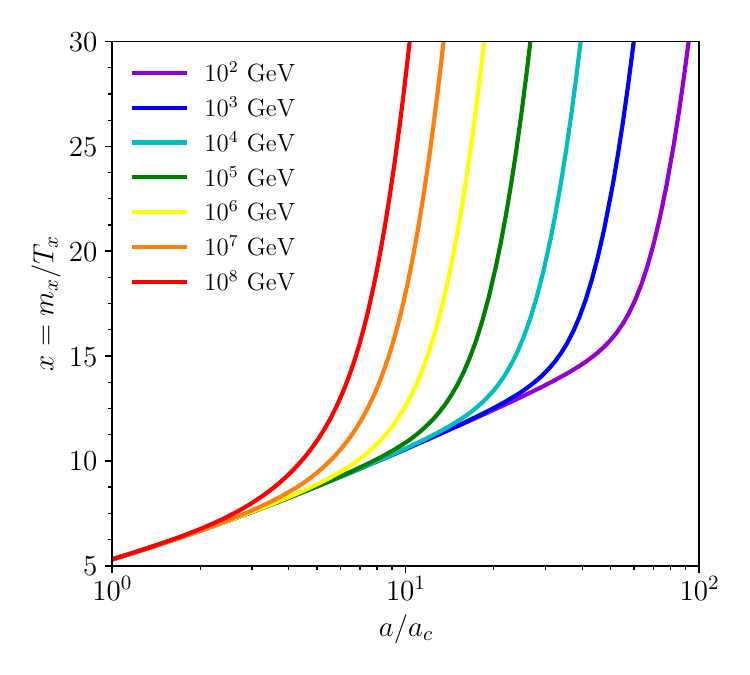}
    \includegraphics[width = 0.47\textwidth]{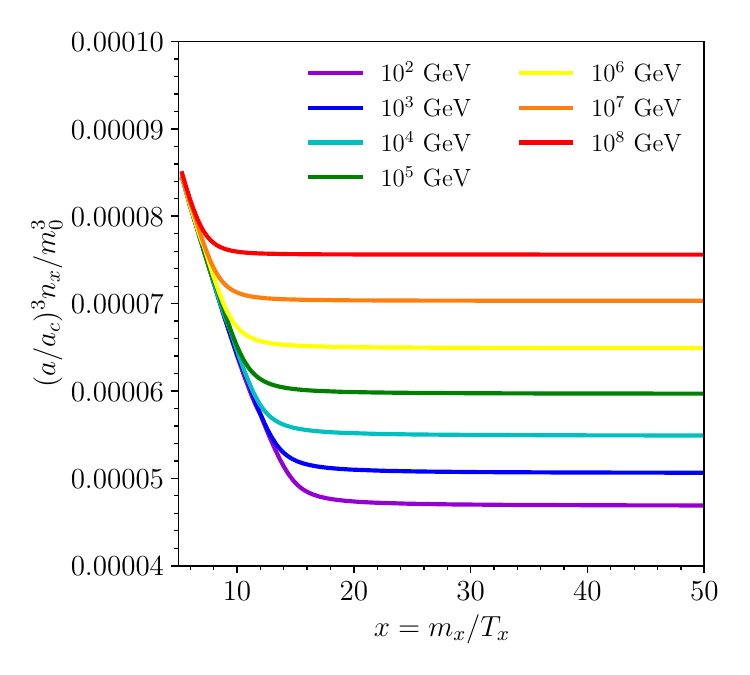}
    \vspace{-0.5cm}
  \end{center}
  \caption{Evolution of the dark glueball temperature $x=m_x/T_x$ with the scale factor relative to confinement~(left), and the rescaled total dark glueball density $(a/a_c)^3n_x/m_0^3$ with the dark temperature~(right) for $m_x = 10^2,\,10^3,\ldots\,,10^8\,\gev$. For this evolution, the dark sector is assumed to be decoupled from the SM with $M\to \infty$ with a negligible energy density in the SM sector.
  }
  \label{fig:pureglue}
\end{figure}

To illustrate the freezeout of the total glueball density, consider first the limit of no transfer reactions ($M\to \infty$) and zero visible density ($\rho\to 0$). The evolution equations derived above then update previous studies of dark glueball freezeout and yield qualitatively similar results.\footnote{The evolution equations of Eqs.~(\ref{eq:dxdt},\ref{eq:dnxdt}) violate entropy conservation in the dark sector during freezeout, which had been assumed in many previous works. Numerically, the impact of this effect is small.} In the left panel of Fig.~\ref{fig:pureglue} we show the evolution of the dark glueball temperature parameter $x=m_0/T_x$ with scale factor $a/a_c$ relative to confinement for several values of $m_0$. Early on, the energy released by $3\to 2$ reactions slows the cooling of the dark plasma to logarithmic in scale factor, an effect called \emph{cannibalism} in Ref.~\cite{Carlson:1992fn}. This continues until $3\to 2$ annihilation freezes out, at which point the dark temperature begins to scale quadratically with the scale factor. In the right panel we show the evolution of the normalized total glueball number density $(a/a_c)^3n_x/m_0^3$ with the temperature parameter $x$. The dark glueball density exhibits a typical freezeout behaviour, falling quickly initially until the annihilation reactions fall out of equilibrium and the yield stabilizes at a nearly constant value. We note as well that for $x\gtrsim 10$, the total dark glueball density is essentially dominated by the $\zpp$ state.

\begin{figure}[ttt]
  \begin{center}
  \includegraphics[width = 0.47\textwidth]{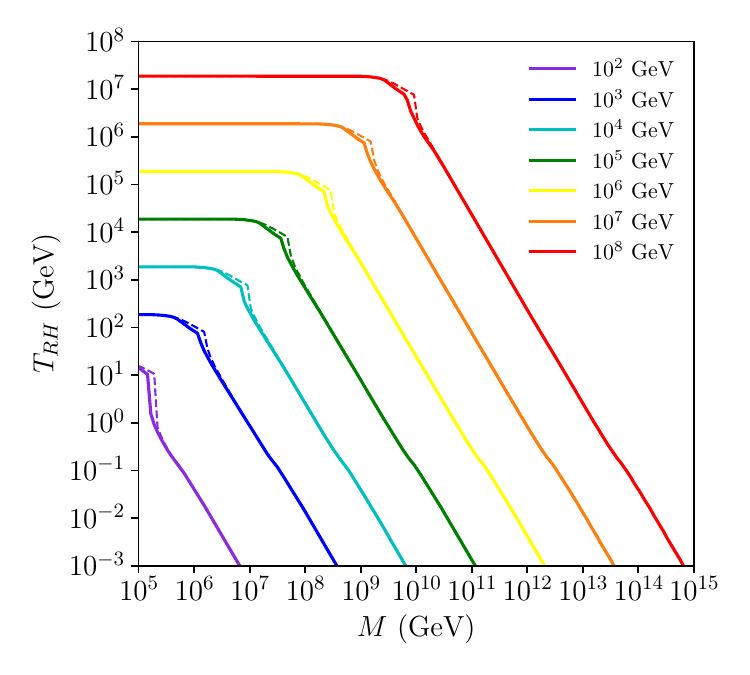}
    \includegraphics[width = 0.47\textwidth]{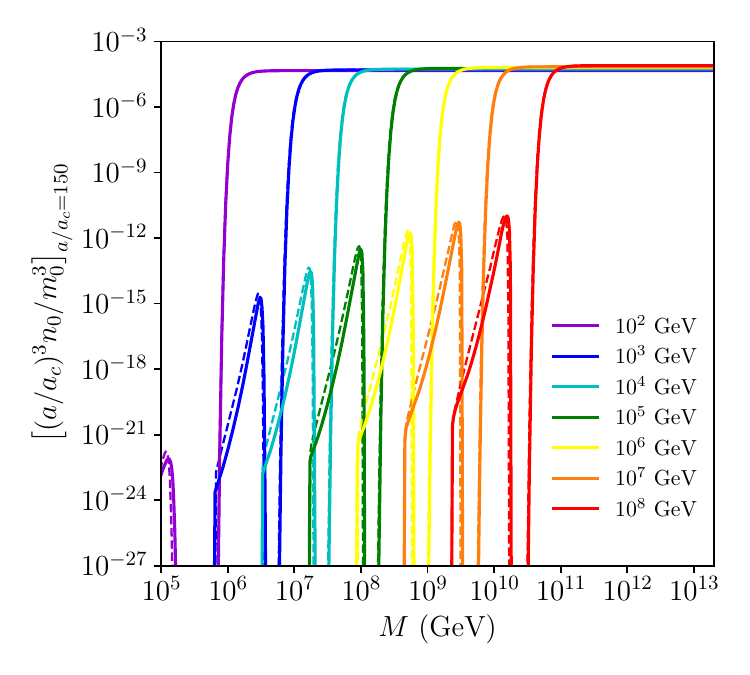}
    \vspace{-0.5cm}
  \end{center}
  \caption{Effect of $\zpp$ decays on the reheating temperature of the visible sector~(left) and the $\zpp$ glueball density~(right) as a function of the connector scale $M$. In the left panel we show the SM reheating temperature after $\zpp$ glueball decays as a function of the connector scale $M$ for $m_0 = 10^2,\,10^3,\,\ldots\,10^8\,\gev$. In the right panel we show the rescaled $\zpp$ glueball number density $(a/a_c)^3n_0/m_0^3$ at the scale factor relative to confinement $a/a_c=150$ for the same values of $m_0$ as a function of $M$. In both panels, the solid lines assume the initial condition $(\rho/\rho_x)_i=0.01$ at $T_{x,i}=5\,T_c$ while the dashed lines have the initial condition $(\rho/\rho_x)_i=g_*/g_{*,x}$.
  }
  \label{fig:gluedec1}
\end{figure}

Let us now return to the full evolution of the dark glueball density and temperature, including decays and a visible component of the total energy density. Since (most of) the dark glueballs must decay to the visible sector to reproduce the observed cosmology, the main physical observable of interest is the reheating temperature after glueball decay, $T_{RH}$. Our working definition for this is the SM sector temperature when $\rho=10\rho_x$ following $\zpp$ decay. In the left panel of Fig.~\ref{fig:gluedec1} we show $T_{RH}$ as a function of the transfer operator mass scale $M$ for several values of $m_0$. The solid lines correspond to the initial condition $(\rho/\rho_x)_i = 0.01$ at the initial dark temperature $T_{x,i}=5\,T_c$ and the dashed lines have $(\rho/\rho_x)_i = g_*/g_{*,x}$ at this starting point. For all curves we see two primary regimes. At smaller $M$, the two sectors are able to equilibrate at the confinement temperature and we obtain $T_{RH} = T_c$. When $M$ is larger, decays of the $\zpp$ glueball begin during or after freezeout of the dark glueball density. For decays well after freezeout we can estimate the reheating temperature by equating $\Gamma_x\simeq H_{RH}$ with $H_{RH} \simeq \sqrt{\rho_{RH}/3\mpl^2}$ and $\rho_{RH}=\pi^2g_*T^4_{RH}/30$ to give
\beq
T_{RH} \ \simeq \ 
0.2\,\gev\lrf{100}{g_*}^{1/4}\!\lrf{m_x}{10^3\,\gev}^{5/2}\!
\lrf{10^8\,\gev}{M}^2 \ .
\eeq
This simple estimate matches quite well with our numerical results and illustrates the scaling $T_{RH}\propto 1/M^2$ for large $M$. 

We also see in Fig.~\ref{fig:gluedec1} that the reheating temperature is typically very similar for the two initial conditions. When $M$ is small, the two sectors thermalize easily for both initial conditions. At larger $M$, the dark glueball sector dilutes more slowly than the SM and comes to dominate the total energy density well before reheating. It is only when $\zpp$ glueball decays occur mainly during freezeout that a non-negligible difference arises. To avoid this early glueball dominance, the initial dark sector energy must be much smaller than the visible sector~\cite{Forestell:2017wov}.

It is also instructive to look at the evolution of the total glueball density when $\zpp$ decays are fast enough to occur immediately after confinement or during the freezeout process. In the right panel of Fig.~\ref{fig:gluedec1} we show the normalized glueball density $(a/a_c)^3n_0/m_0^3$ at the scale factor $(a/a_c)= 150$ relative to confinement. This particular value is reasonably close to confinement while being large enough to fall after dark glueball freezeout in the absence of decays (as seen in Fig.~\ref{fig:pureglue}). The solid and dashed lines correspond to the same two initial conditions discussed above. For larger $M$, the $\zpp$ decays occur for $a > 150\,a_c$ and do not affect the glueball relic density at this point. As $M$ decreases, decays become faster and eventually the glueball density at $(a/a_c)=150$ falls quickly. However, as $M$ decreases further the normalized density rises before falling off again. 

This rebound in density comes from a net heating of the glueball ensemble through decays, which can be large enough to cause a modest decrease in $x$ (increase in $T_x$) in some regimes. Specifically, from Eqs.~(\ref{eq:cntr},\ref{eq:cetr}) there is a preference for faster-moving glueballs to decay more slowly due to time dilation. Thus, decays remove slower glueballs preferentially leaving behind glueballs with more energy per particle on average. If the decays are fast enough, this self-heating effect, combined with a slowdown in cooling from the decoupling of heavier glueballs, can overcome the cooling from expansionary dilution. In turn, this decreases how quickly the total glueball density falls off leading to larger remaining densities at the reference scale factor $(a/a_c)=150$ than would occur without the decay heating effect. 

The heating effect from decays seen here can be countered to some extent by energy transfer with the visible sector through the elastic scattering of dark glueballs with particles in the SM plasma mediated by the dimension-6 operator of Eq.~\eqref{eq:ltr}. We have examined this scattering following the methods of Refs.~\cite{Bringmann:2006mu,Kuflik:2017iqs} and have incorporated it into the evolution equations. However, we find that the rate of energy transfer from elastic scattering is suppressed relative to decays by several factors of $T_x/m_0,\,T/m_0 \ll 1$. Elastic scattering in this context turns out to be too weak to significantly counter heating from decays.

\subsection{Evolution of the $\opm$ Glueball Density}

We turn next to the freezeout of the heavier $\opm$ glueball state. This process is dominated by the $2\to 2$ annihilation reactions $\opm+\opm \to \zpp+\zpp$. When the $\zpp$ is long-lived relative to freezeout of this reaction, this process occurs in the background of a massive bath dominated by $\zpp$ glueballs, as studied previously in a general context in Ref.~\cite{Pappadopulo:2016pkp,Farina:2016llk} and specifically for glueballs in Refs.~\cite{Forestell:2016qhc,Forestell:2017wov}.

The evolution of the $\opm$ density is described by
\beq
\frac{dn_1}{dt} = -3Hn_1 - \mathcal{C}_1^{x} \ ,
\eeq
with $\mathcal{C}_1^{x}$ given by Eq.~\eqref{eq:c1x}.\footnote{For now we neglect potential decays of the $\opm$ glueball.
} 
Relative to standard freezeout in a thermalized relativistic bath, the primary difference with a massive massive thermal bath comes from the backreaction term in the collision operator. When annihilation reactions are fast, the number density is driven to
\beq
n_1 \ \to \ \frac{n_0}{\bar{n}_0(x)}\,\bar{n}_1 
\ \simeq \ \frac{n_x}{\bar{n}_x(x)}\,\bar{n}_1 \ ,
\eeq
where $\bar{n}_i(x)$ is the equilibrium density at $x=m_0/T_x$. The prefactor $n_x/\bar{n}_x$ is significantly enhanced relative to unity after freezeout of the total glueball number density. In this way, the massive bath can support a larger density of the $\opm$ mode through backreactions. 

\begin{figure}[!ttt]
  \begin{center}
    \includegraphics[width = 0.47\textwidth]{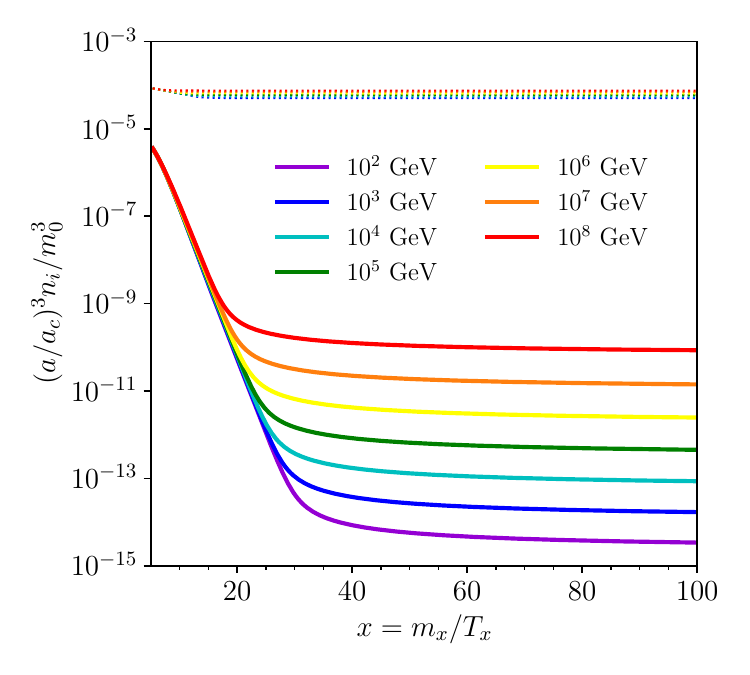}
    \includegraphics[width = 0.47\textwidth]{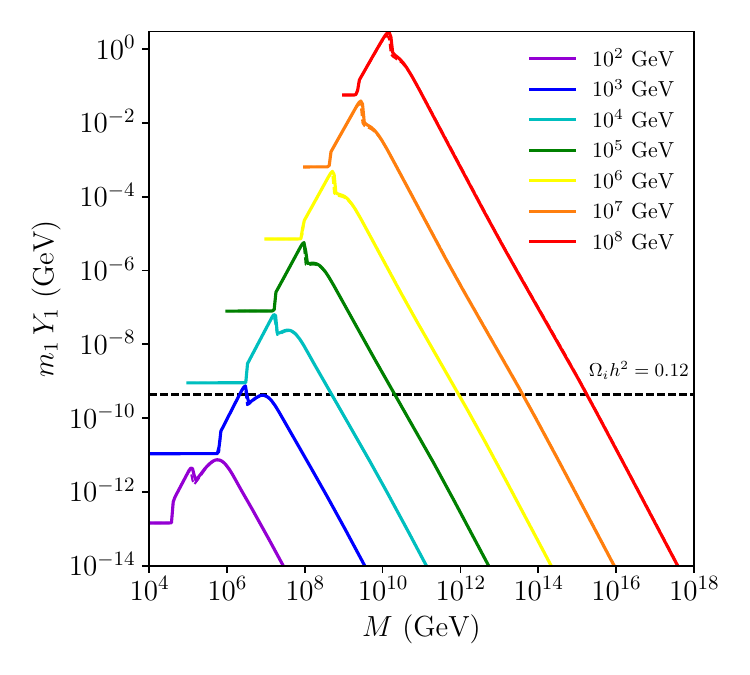}
    \vspace{-0.5cm}
  \end{center}
  \caption{Relic densities of the $1^{\pm}$ glueball 
without~(left) and with~(right) $\zpp$ decays. In the {left} panel we show the evolution of the rescaled densities $(a/a_c)^3n_i/m_0^3$ of the $\opm$~(solid) and total glueball~(dashed) as functions of $x=m_0/T_x$ for $m_0=10^2,\,\ldots,\,10^8\,\gev$. In the {right} panel we show the asymptotic ($t\to \infty$) value of the mass-weighted yield $m_1Y_1$ of the $\opm$ glueball when decays are included as a function of the connector scale $M$ for the same set of sample $m_0$ values.
  }
  \label{fig:glu1a}
\end{figure}

To illustrate the evolution of the $\opm$ density in a massive background bath, we show in the left panel of Fig.~\ref{fig:glu1a} the total~(dashed) and $\opm$~(solid) normalized glueball densities $(a/a_c)^3n_i/m_0^3$ as functions of the dark temperature parameter $x$ in the limit of no glueball decays ($M\to \infty$) and no SM population for a range of values of $m_0$. Freezeout of the $\opm$ state occurs significantly later than the freezeout of the total glueball density, justifying our previous assumption of $n_i/\bar{n}_i=n_0/\bar{n}_0$. The resulting $\opm$ density is also much smaller than the total relic glueball density, but larger than it would have been for freezeout into a relativistic bath.

Returning to the full scenario with finite $M$, we show in the right panel of Fig.~\ref{fig:glu1a} the asymptotic ($a\to \infty$) value of the mass-weighted yield $m_1Y_1$ of the $\opm$ glueball for several values of $m_0$ as a function of $M$, where $Y_1=n_1/s$ is normalized to the total entropy of the visible sector after reheating. The initial condition is taken to be $(\rho/\rho_x)_i=0.01$ but the results for $(\rho/\rho_x)_i=g_*/g_{*,x}$ are nearly indistinguishable due to early glueball dominance or thermalization for both initial conditions as discussed above. When $\zpp$ decays are fast at confinement, corresponding to smaller $M$, the massive glueball background disappears nearly instantaneously and the final $\opm$ yield goes to a constant value independent of $M$. Going to slightly larger $M$, the $\opm$ relic density increases since now the decays occur a bit later during the freezeout process and there is a larger background to support the $\opm$ density through inverse reactions leading to later freezeout. In this rising region we also see a feature due to the heating effect of decays discussed previously. Increasing $M$ further then leads to a turnover and decrease in the final yield. This corresponds to $\zpp$ decays well after both the total glueball and $\opm$ densities have frozen out. Prior to the $\zpp$ decay, the ratio $(n_1/n_x)_{fo}$ from freezeout is independent of $M$ in this regime. However, the later the $\zpp$ glueballs decay the greater the subsequent dilution of the $\opm$ density. A simple analytic estimate gives $m_1Y_1 \sim T_{RH}(n_1/n_x)_{fo}$, which scales as $1/M^2$ in agreement with our numerical result. Note that we truncate all the curves in Fig.~\ref{fig:glu1a} at $m_0/M = 1/10$ where we expect the effective description of the transfer operator in Eq.~\eqref{eq:ltr} to become unreliable.

\section{Cosmological Implications\label{sec:cosmo}}
We have shown that a dominant initial dark gluon sector in the early universe can confine, decay, and reheat the visible sector. A necessary remnant of this process is a relic density of long-lived or stable $\opm$ glueballs. In this section we investigate the cosmological implications of the scenario for two distinct cases. In the first, we assume that dark charge conjugation number $C_x$ is conserved implying that the $\opm$ glueball is stable and a component of DM. For the second case we consider broken $C_x$ and decays of the $\opm$ glueballs through the dimension-8 operator of Eq.~\eqref{eq:nocx}. In both cases, we study the impact on cosmological and laboratory observables.

\subsection{Stable $\opm$ Glueballs as Dark Matter}

\begin{figure}[!ttt]
  \begin{center}
    \includegraphics[width = 0.85\textwidth]{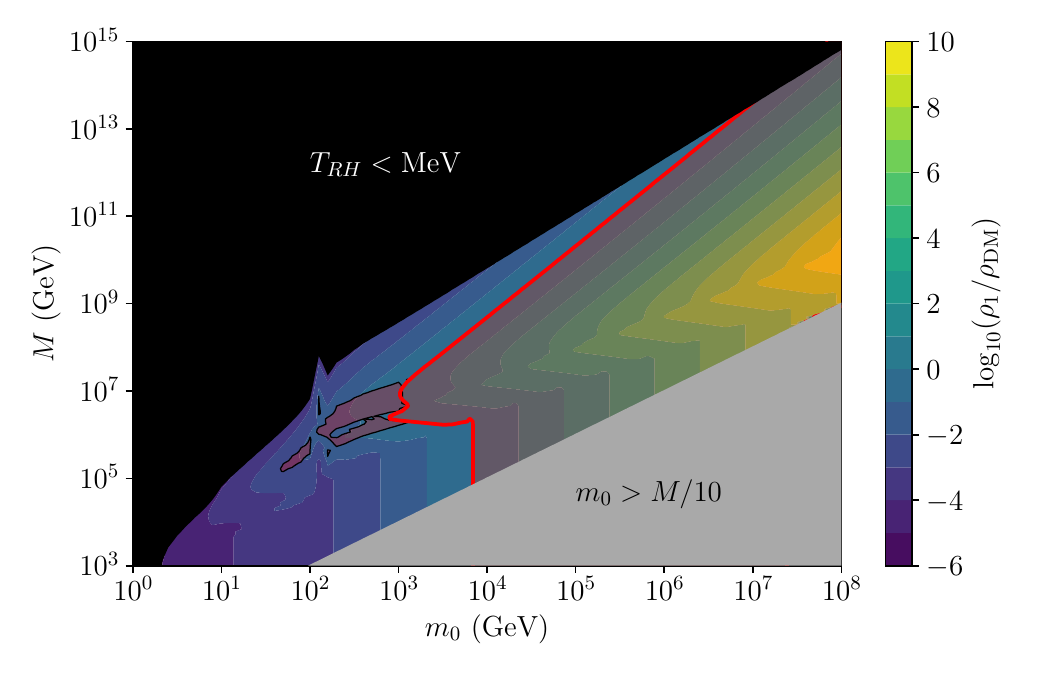}
            \vspace{-0.5cm}
  \end{center}
  \caption{Relic density of a stable $\opm$ glueball relative to the observed DM density in the $m_0$--$M$ plane. The solid red line indicates where the $\opm$ glueball can make up the full DM abundance. The shaded region to the right of this line is ruled out by dark glueball overclosure. The red shaded region to the left of the line is excluded by cosmic ray signals from residual glueball annihilation. The upper left black region has $T_{RH} < \mev$ and is ruled out by BBN, while the lower right grey region has $m_0 > M/10$ where the effective connector operator approach used in our calculation is expected to be unreliable.
  }
  \label{fig:m1y1}
\end{figure}

If dark charge conjugation number $C_x$ is conserved, the lightest $C_x$-odd glueball, the $\opm$ state, is stable. It will therefore contribute to the density of DM. In Fig.~\ref{fig:m1y1} we show the relic density of the $\opm$ glueball relative to the observed DM energy density in the $m_0$--$M$ plane assuming a subleading initial SM energy fraction $(\rho/\rho_x)_i=0.01$ at an initial dark sector temperature of $T_{x,i} = 5T_c$. The $\opm$ relic density for the second thermalized initial condition $(\rho/\rho_x)_i=g_*/g_{*,x}$ is nearly identical. In addition to the relic density, we also block off the upper left region in which the SM reheating temperature after $\zpp$ glueball decay lies below $T_{RH} < 1\,\mev$ and is ruled out by primordial nucleosynthesis~\cite{Hannestad:2004px,deSalas:2015glj}. To the lower right we block the region where $m_0 > M/10$ where our effective theory description of the transfer operators in Eq.~\eqref{eq:ltr} is expected to become unreliable.

The relic $\opm$ density depends on both $m_0$ that determines the glueball masses and $M$ that modulates the strength of connection between the dark and visible sectors. Depending on $M$, the $\opm$ state can make up the full DM density for a range of $m_0$ between about $10^3$--$10^7\,\gev$, shown by the solid red contour line. In the red shaded region to the right of the red contour line, the scenario predicts too much $\opm$ glueball DM and is ruled out. 

Relic $\opm$ glueballs can also generate potentially observable signals in indirect detection searches, even when they have an acceptable relic density. Remnant annihilations $\opm+\opm \to \zpp+\zpp$ will produce visible products through the subsequent $\zpp$ decay to the SM. Based on the $\opm$ density and annihilation rates (and $\zpp$ decay final states), we find a non-trivial exclusion from combined gamma ray measurements~\cite{Hess:2021cdp}. This is shown in Fig.~\ref{fig:m1y1} by the shaded region up to the left of the red solid line.

If the $\opm$ comprises an appreciable fraction of the DM density, it can lead to a signal in direct detection experiments via Higgs exchange through the dimension-6 operator of Eq.~\eqref{eq:ltr}. The cross section for this process is
\begin{equation}
\begin{aligned}
\sigma&\sim 10^{-49}~{\rm cm}^2\left(\frac{1~\rm TeV}{m_0}\right)^2\left(\frac{m_1/m_0}{1.78}\right)^2\left(\frac{10}{M/m_0}\right)^4\left(\frac{3}{N}\right)^2 \ .
\end{aligned}
\label{eq:sigmaSI}
\end{equation}
Combining this cross section with the predicted $\opm$ relic density, we find that the signal is too small to be observed by current or near future experiments.

Other potential tests of this scenario include collider searches, particularly through exotic Higgs decays~\cite{Curtin:2023skh,Batz:2023zef,Bishara:2024rtp}, and modifications to cosmic structure formation~\cite{Erickcek:2020wzd,Erickcek:2021fsu,Barenboim:2021swl}. Over the parameter space considered we do not find any current exclusions from these probes.

So far we have concentrated on glueballs from the dark gauge group $G_x=SU(3)$. Our results can be generalized to $SU(N > 3)$ and some other non-Abelian groups. Increasing $N$ is expected to only produce small modifications to the confinement temperature and glueball spectrum~\cite{Teper:1998kw,Lucini:2012gg}. Our large-$N$ estimates also suggest that this will not significantly impact the total $\zpp$ decay rate. However, these same estimates indicate that glueball annihilation will be significantly suppressed for larger $N$, with cross sections scaling like $\langle\sigma_{32}v^2\rangle \propto (3/N)^6$ and $\langle\sigma_{22}v^2\rangle \propto (3/N)^4$ from Eqs.~(\ref{eq:sig32},\ref{eq:sig22}). Together, these arguments indicate that for larger $N$ the visible reheating temperature will be nearly the same for given values of $m_0$ and $M$ while the $\opm$ relic density will increase. In particular, the region where the $\opm$ generates the full DM abundance will be shifted to lower glueball masses.

\subsection{Decays of an Unstable $\opm$ Glueball}

If dark sector charge conjugation number $C_x$ is broken, the dimension-8 transfer operator in Eq.~\eqref{eq:ltr} is allowed and the $\opm$ state can decay with a rate that is parametrically longer than the other glueballs. This implies that the $\opm$ will decay after the visible sector has been reheated by $\zpp$ decays. Depending on when the $\opm$ decays take place and how abundant this state was to begin with, these decays can disrupt BBN or the CMB, or create observable cosmic rays today.

\begin{figure}[ttt]
  \begin{center}
    \includegraphics[width = 0.85\textwidth]{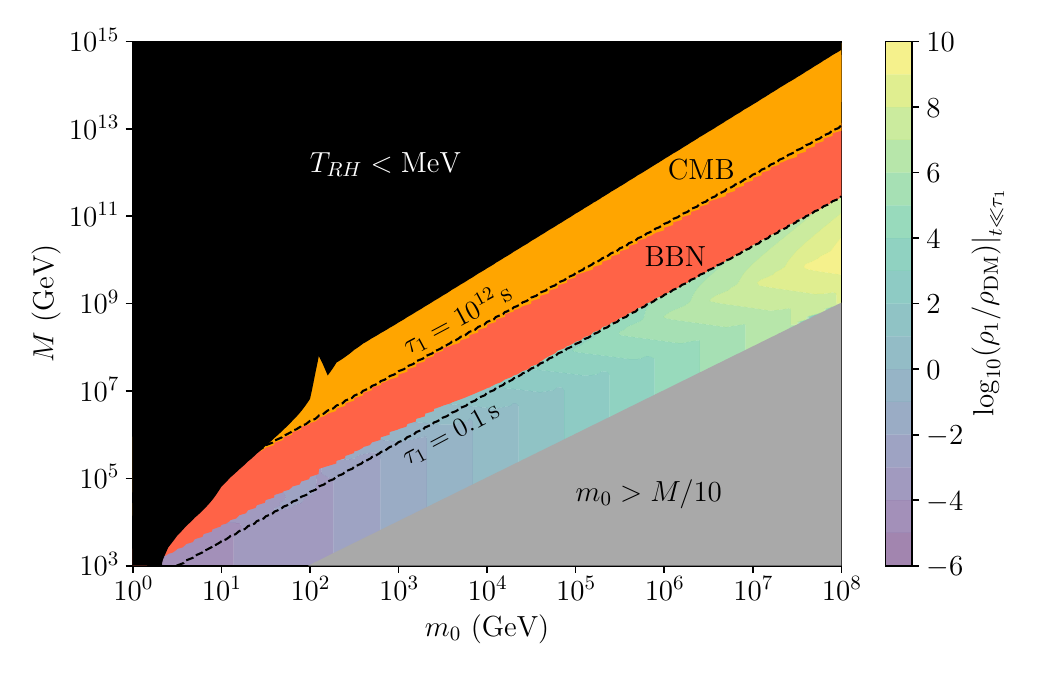}
    \vspace{-1cm}
  \end{center}
  \caption{Cosmological bounds in the $m_0$--$M$ plane for the scenario with broken $C_x$ and decays of the $\opm$ glueball in the early universe through the dimension-8 operator of Eq.~\eqref{eq:ltr}. Decays of relic $\opm$ glueballs are ruled out by the observed CMB power spectrum in the upper orange region and by BBN in the red region below. The upper left black region has $T_{RH} < \mev$ and cannot lead to successful BBN, while the lower right grey region has $m_0 > M/10$ where the effective operator treatment of connectors is expected to be unreliable. Below the BBN exclusion, relic $\opm$ glueballs would have decayed well before today and this is region is consistent with existing observations. For reference, in this region we display contours of the $\opm$ energy density before it decays. The lower and upper black dashed lines indicate $\opm$ glueball lifetimes of $\tau_1 = 0.1,\,10^{12}\,\text{s}$.
  }
  \label{fig:m1y1dec}
\end{figure}

In Fig.~\ref{fig:m1y1dec} we show current bounds on the scenario with decaying $\opm$ glueballs in the $m_0$--$M$ plane. The dark red shaded region in Fig.~\ref{fig:m1y1dec} is excluded by energy injection from glueball decays during BBN, based on the results of Refs.~\cite{Kawasaki:2015yya,Kawasaki:2017bqm}. The lower boundary of this exclusion region corresponds to lifetimes $\tau_1 \simeq 0.1\,\text{s}$, shown by the lower dashed black line, which is roughly when BBN begins. Bounds from energy injection into the CMB and the impact it has on the power spectrum are shown by the orange shaded region in the figure, and are estimated using the results of Refs.~\cite{Slatyer:2016qyl,Yang:2015cva}. The lower boundary of this exclusion coincides approximately with $\tau_1 \simeq 10^{12}\,\text{s}$, indicated by the upper black dashed line, corresponding to the time of recombination. As before, we also show in Fig.~\ref{fig:gluedec1} exclusions from $T_{RH} < \mev$ after $\zpp$ decays in the upper left and $m_0 > M/10$ at the lower right. 

In the region below the BBN exclusion but above the lower grey region, all the relic $\opm$ glueballs will have decayed well before today and the scenario is consistent with current experimental and observational bounds. For reference, in this region we superimpose contours of the $\opm$ relic density \emph{before it decays}. We also note that in some parts of this allowed region the $\opm$ relic density can be large enough to come to dominate the total energy density of the universe before it decays. This implies yet another stage of reheating of the visible (necessarily to temperatures above a few MeV) and an additional dilution of any other pre-existing relics.

As a slight variation on this decaying $\opm$ scenario, one can also consider the case of $C_x$ breaking exclusively by new physics near the Planck scale $\mpl$ but $C_x$-conserving connectors to the SM at a much lower scale $M$. This picture would correspond approximately to maintaining the dimension-6 connector operator in Eq.~\eqref{eq:ltr} with $M \ll \mpl$ but replacing $M\to \mpl$ in the dimension-8 operator. We find that with this replacement the $\opm$ glueball has a lifetime many orders of magnitude larger than the age of the universe and even the decay lifetimes probed by searches for decaying DM. The cosmological implications of this related scenario are therefore identical to the case of the $\opm$ glueball being exactly stable.

\bigskip

\section{Conclusions\label{sec:conc}}

In this work we have studied the cosmological implications of a minimal non-Abelian dark gauge sector when it is populated preferentially by primordial inflation. Our focus has been on a pure Yang-Mills dark sector based on the gauge group $\sun$ (with $N=3$ for the most part). We find that this simple scenario can be consistent with observations and provide a viable DM candidate in the form of a glueball over a broad range of masses.

The story of cosmological evolution and DM in this theory is dominated by two states: the lightest glueball in the spectrum with $\jpc$ quantum numbers $\zpp$, and lightest $C$-odd glueball with quantum numbers $\opm$. Our assumed starting point is a hot bath of dark gluons created by primordial inflation with temperature $T_x$ above the dark confinement scale $\Lambda_x$. Once created, the thermal bath of dark gluons expands, cools, and eventually confines when the temperature approaches the confinement scale, $T_c \sim \Lambda_x$. The dark sector below the confinement temperature consists of a bath of glueballs. These undergo a freezeout process that yields a dominant population of the lightest $\zpp$ glueball in the spectrum and much smaller densities of the heavier states. To reproduce the observed cosmology, the $\zpp$ glueball must decay to populate and reheat the visible sector of SM particles.

Decays of the $\zpp$ glueball to the SM can arise at lowest order from the dimension-6 Higgs portal operator given in Eq.~\eqref{eq:ltr}. This operator (and its $PT$ conjugate) also mediate decays of all the other glueballs save one, the lighest $C$-odd $\opm$ glueball. If $C$ is conserved in the dark sector, the $\opm$ is stable and will contribute to the total density of DM. When $C$ is broken in the dark sector, the dimension-8 operator of Eq.~\eqref{eq:ltr} can mediate the decays $\opm \to \zpp + Z/\gamma$. In this latter case, since the decay operator is of higher dimension, the $\opm$ will be parametrically long-lived relative to the other glueballs. Its decays after reheating by the $\zpp$ mode can impact cosmological observable such as BBN and the CMB and generate high-energy cosmic rays.

In our study we have investigated in detail the cosmological evolution of the dark and visible sectors, including the effects of glueball self-interactions, transfer reactions between the sectors, and dynamics of the confining phase transition based on results from lattice calculations. Our work builds upon and refines previous studies of dark glueball dynamics. We focus on the scenario of early dark gluon dominance, as opposed to cases where the initial dark sector density is a sub-leading component, and we study in greater detail the relic density of $\opm$ glueballs obtained in this scenario. By including the full lattice-derived equation of state of the dark glueball sector prior to confinement, we improve the determination of energy transfer before confinement. We have also included decay reheating and freeze out effects involving the entire glueball spectrum more completely than before. While our updated approach agrees qualitatively with previous work, some of the detailed quantitative results are different. 

Our study has implications for new dark gauge sectors beyond the SM. When dark sector $C_x$ is conserved, this scenario yields a viable DM candidate from the $\opm$ glueball for a range of glueball masses $m$ and connector operator scales $M$. However, over a wide range of parameters the visible reheating temperature after $\zpp$ decay is unacceptably low or the relic density of $\opm$ glueballs is too high (or ruled out by indirect searches for DM). This implies broad constraints on the existence of new dark gauge forces, and it limits what is possible for such dark sectors to produce potentially observable signals in gravitational waves from a first-order dark confining phase transition. When dark sector $C_x$ is broken, the relic population of $\opm$ glueballs can decay and modify BBN or the CMB, yielding a complementary set of constraints within this second scenario.

\begin{flushleft}
\textit{Note added:} as this work was nearing completion the study of Ref.~\cite{Biondini:2024cpf} appeared that investigates a closely related scenario of dark glueball evolution.
\end{flushleft}

\section*{Acknowledgements}
We thank Carlos de Lima, Navin McGinnis, and Douglas Tuckler
for helpful comments and discussions.
This work is supported in part by the Natural Sciences and Engineering Research Council of Canada~(NSERC). TRIUMF receives federal funding via a contribution agreement with the National Research Council~(NRC) of Canada.


\bibliography{refs}

\end{document}